\documentclass{aastex631}

\usepackage{amsmath}
\usepackage{orcidlink}
\usepackage{subfigure}
\pdfoutput=1
\definecolor{ForestGreen}{RGB}{34,139,34}
\shorttitle{}
\shortauthors{}
\graphicspath{{./}{figures/}}

\begin{document}

\title{Benchmarking Exact, GP-Emulated, and Simulation-Based Inference for Late-Time Cosmology}

\author[0000-0003-0580-0798]{Sai Swagat Mishra}\thanks{saiswagat009@gmail.com}
\affiliation{SR University, Warangal, Telangana, 506371}



\begin{abstract}

Forthcoming cosmological surveys require inference pipelines that are both statistically reliable and computationally scalable. In this work, we perform a systematic comparison of three complementary inference strategies for late-time $\Lambda$CDM cosmology: exact Markov Chain Monte Carlo (MCMC), Gaussian Process (GP)-assisted MCMC, and neural Simulation-Based Inference (SBI). Using a common analysis framework based on Cosmic Chronometers, DESI DR2 baryon acoustic oscillation measurements, and the Pantheon+ Type Ia supernova compilation, we consider two dataset combinations of increasing complexity, namely CC+DESI and CC+DESI+PP, under identical cosmological assumptions and priors. For CC+DESI, both GP emulation and SBI reproduce the exact posterior constraints on $(H_0,\Omega_{m,0})$ to better than $0.3\sigma$. For the more constraining CC+DESI+PP combination, modest method-dependent shifts emerge, reaching at most $\sim1.5\sigma$ in a single parameter. Despite these differences, all methods recover a nearly identical expansion history, with percent-level agreement across the full redshift range. From a computational perspective, GP emulation accelerates model evaluations but remains limited by MCMC sampling, whereas SBI achieves order-of-magnitude reductions in total runtime through amortized posterior learning. We further investigate the convergence of SBI as a function of simulation budget and identify the number of simulations required to obtain stable posterior constraints. Overall, our results demonstrate that accelerated inference techniques can deliver reliable cosmological constraints for realistic late-time datasets considered here at a fraction of the computational cost of conventional likelihood-based analyses.

\end{abstract}



\section{Introduction}

\label{sec:intro}

The $\Lambda$CDM model has emerged as the standard framework for describing the evolution of the Universe, providing a remarkably successful description of cosmological observations over a wide range of scales and epochs. Measurements of the cosmic microwave background (CMB), large-scale structure, baryon acoustic oscillations (BAO), Type Ia supernovae (SN Ia), and direct probes of the expansion history have collectively established a cosmological model dominated by cold dark matter and a cosmological constant \citep{SupernovaSearchTeam:1998fmf, SupernovaCosmologyProject:1998vns,SDSS:2005xqv,Weinberg:2013agg,Planck2018}. Within this framework, a small number of cosmological parameters accurately describes the observed expansion history and growth of structure, enabling cosmology to enter an era of precision science.

At the same time, steadily improving observations have exposed a number of challenges that place increasing demands on cosmological parameter estimation. Among the most notable are the persistent discrepancies in the inferred value of the Hubble constant $H_0$ and the amplitude of matter clustering, often referred to as the $H_0$ and $S_8$ tensions \citep{Planck2018,Riess2022SH0ES,Verde2019Tension, Mishra:2025rhi}. Whether these tensions signal unaccounted systematic effects, limitations of the standard cosmological model, or hints of new physics remains an open question. Addressing these issues requires not only increasingly precise observations but also robust and reliable statistical inference techniques capable of extracting the maximum amount of information from modern cosmological datasets.

Late-time cosmological probes play a particularly important role in this context. Cosmic Chronometers (CC) provide direct measurements of the Hubble expansion rate through the differential-age technique, BAO observations constrain the geometric evolution of the Universe through the characteristic sound horizon scale imprinted in the matter distribution, and Type Ia supernovae trace the luminosity-distance relation over a broad redshift range. Together, these probes provide complementary and largely independent constraints on the expansion history, the present-day matter density parameter $\Omega_{m,0}$, and the Hubble constant $H_0$. The combination of these datasets has therefore become a cornerstone of modern low-redshift cosmology and an important testing ground for both the standard cosmological model and its possible extensions.

As observational datasets continue to grow in size and precision, the computational cost of cosmological inference has become an increasingly important challenge. Exact Markov Chain Monte Carlo (MCMC) methods remain the benchmark for Bayesian parameter estimation owing to their statistical rigor and asymptotic exactness \citep{MCMC_review,Speagle:2019ffr}. However, the increasing dimensionality of modern datasets and the complexity of their associated likelihood functions have motivated the development of accelerated inference techniques. Broadly speaking, these approaches either replace expensive theoretical calculations with surrogate models, as in Gaussian-process (GP) emulation \citep{GP_emulator_cosmo2,Lawrence:2017ost,Bocquet:2020tes}, or learn posterior distributions directly from simulations through simulation-based inference (SBI) using neural density estimators \citep{SBI_review,Wang:2023vej}. Both strategies offer the potential for substantial reductions in computational cost while preserving the statistical information contained in the data.

Despite rapid progress in emulator-based and simulation-based approaches, relatively few studies have performed controlled, like-for-like comparisons of these methodologies using realistic late-time cosmological datasets. In particular, the practical trade-off between computational efficiency and statistical fidelity remains insufficiently quantified. While exact MCMC provides a natural benchmark, it is not always clear how closely approximate approaches reproduce the resulting posterior distributions, how any deviations depend on the constraining power of the data, and how the computational gains scale with the complexity of the likelihood.

In this work, we address these questions within the framework of late-time $\Lambda$CDM cosmology. We consider three complementary low-redshift probes: CC, anisotropic BAO measurements from the Dark Energy Spectroscopic Instrument (DESI) Data Release 2, and Type Ia supernovae from the Pantheon+ compilation. From these datasets we construct two combinations of increasing complexity and constraining power: CC+DESI and CC+DESI+PP, where PP denotes Pantheon+. The former provides competitive constraints on $(H_0,\Omega_{m,0})$ using direct expansion-rate and geometric information, while the latter combines thousands of measurements into a substantially higher-dimensional likelihood with significantly greater computational demands.

Using these datasets, we perform a systematic comparison of three distinct inference strategies: exact MCMC sampling of the full likelihood, GP-assisted MCMC based on surrogate modelling of cosmological observables, and simulation-based inference implemented through neural posterior estimation. All three methods employ identical cosmological models, priors, and observational datasets, enabling a direct comparison of parameter constraints, posterior distributions, convergence behavior, and computational efficiency.

The primary goals of this work are fourfold. First, we assess the statistical fidelity of GP emulation and SBI by comparing the recovered posterior distributions and cosmological parameter constraints against those obtained from exact MCMC. Second, we examine the extent to which the different inference schemes recover the same underlying cosmological evolution by reconstructing the Hubble expansion history and quantifying deviations from the exact-likelihood benchmark. Third, we quantify the computational cost of each approach and identify regimes in which accelerated inference techniques provide significant advantages over conventional likelihood-based sampling. Finally, we investigate the convergence properties of SBI as a function of simulation budget, thereby determining the number of simulations required to achieve stable posterior constraints and agreement with the exact posterior at a statistically meaningful level.

This manuscript is organized as follows. In Section~\ref{sec:methods_data}, we describe the inference methodologies, datasets, and implementation details employed in the analysis. Section~\ref{sec:results} presents the cosmological constraints, reconstructed expansion history, runtime comparisons, and SBI convergence studies. Finally, Section~\ref{sec:conc} summarizes our main findings and discusses future directions for accelerated cosmological inference.
\newpage

\section{Methods and data}
\label{sec:methods_data}
\subsection{Cosmological model}

Throughout this work, we assume a spatially flat $\Lambda$CDM cosmology, in which the late-time expansion of the Universe is driven by pressureless matter and a cosmological constant. Neglecting the contribution of radiation at the redshifts considered here, the Hubble parameter is given by

\begin{equation}
H(z)=H_0E(z),
\end{equation}

where the dimensionless expansion rate is

\begin{equation}
E(z)=\sqrt{\Omega_{m,0}(1+z)^3+(1-\Omega_{m,0})}.
\end{equation}

Here, $H_0$ denotes the present-day Hubble constant and $\Omega_{m,0}$ is the present matter density parameter. Consequently, the parameter vector constrained in this analysis is

\begin{equation}
\boldsymbol{\theta}=(H_0,\Omega_{m,0}).
\end{equation}

For a given set of cosmological parameters, the corresponding observables entering the CC, DESI, and PP likelihoods are computed using the above expansion history. These theoretical predictions form the basis of the Exact MCMC, GP-emulated, and SBI analyses presented in this work. In all analyses, we adopt flat priors
$60 < H_0 < 80~{\rm km\,s^{-1}\,Mpc^{-1}}$,
$0.2 < \Omega_{m,0} < 0.4$
,
with no external prior imposed on either parameter.

\subsection{Datasets}

To evaluate the performance of the different inference schemes, we employ three complementary late-time cosmological probes: Cosmic Chronometers, anisotropic Baryon Acoustic Oscillation (BAO) measurements from DESI DR2, and Type Ia Supernova observations from the Pantheon+ compilation. These datasets provide independent constraints on the expansion history and distance-redshift relation of the Universe, enabling a robust assessment of both parameter recovery and computational performance.

\paragraph{Cosmic Chronometers (CC).}
Cosmic Chronometers provide direct measurements of the Hubble parameter through the differential-age technique,
\begin{equation}
H(z)=-\frac{1}{1+z}\frac{dz}{dt}.
\end{equation}
Unlike distance-based probes, CC measurements constrain the expansion history directly and are largely model independent. In this work, we employ a compilation of CC measurements spanning the late-time Universe, incorporating both correlated and uncorrelated data points using the corresponding covariance information where available \citep{Jimenez:2001gg, Moresco:2016mzx, Moresco:2020fbm, Moresco:2022phi}.

\paragraph{DESI DR2 BAO.}
We further include anisotropic BAO measurements from the second data release of the Dark Energy Spectroscopic Instrument (DESI DR2). The dataset provides constraints on the transverse and radial distance scales through measurements of $D_M(z)/r_d$ and $D_H(z)/r_d$, where $D_M(z)$ is the transverse comoving distance, $D_H(z)=c/H(z)$, and $r_d$ denotes the sound horizon at the drag epoch. The full covariance matrix supplied by the DESI collaboration is incorporated in the likelihood analysis \citep{DESI:2024mwx, DESI:2025zgx}. A comprehensive discussion on the dataset can be found in \citep{Mishra:2025vpy}.
In the present analysis, the sound horizon at the drag epoch is fixed to
$r_d=147.78~{\rm Mpc}$ \citep{Planck2018} and is not treated as an additional free parameter.
This value is used consistently in the DESI observables
$D_M(z)/r_d$ and $D_H(z)/r_d$ for all three inference schemes.
\paragraph{Pantheon+.}
To further constrain the luminosity-distance relation, we employ the Pantheon+ Type Ia Supernova compilation, consisting of more than 1700 spectroscopically confirmed supernovae spanning the redshift range $0.001\lesssim z \lesssim 2.3$.
 We use the Pantheon+SH0ES-calibrated distance-modulus data vector
provided with the adopted compilation, together with its full covariance
matrix containing both statistical and systematic contributions
\citep{Brout:2022vxf,Scolnic:2021amr}.
The absolute magnitude of the supernovae is therefore not introduced as
an additional free nuisance parameter in our two-parameter inference; rather, the absolute distance calibration is inherited from the adopted
Pantheon+SH0ES data product. No independent Gaussian prior on $H_0$ or
additional SH0ES likelihood is imposed.
For brevity, the Pantheon+ compilation is referred to as PP throughout
this manuscript. 

Throughout this work, we consider two dataset combinations, namely CC+DESI and CC+DESI+PP, allowing us to investigate the performance of the inference methods across likelihoods of significantly different dimensionality and computational complexity. Both CC+DESI and CC+DESI+PP are analyzed independently with Exact
MCMC, GP-assisted MCMC, and SBI.
\subsection{Likelihood construction}

For the combined analyses, the total likelihood is constructed as the
product of the individual likelihoods, or equivalently as the sum of
their $\chi^2$ contributions,
\begin{equation}
\chi^2_{\rm tot}
=
\chi^2_{\rm CC}
+
\chi^2_{\rm DESI}
+
\chi^2_{\rm PP},
\end{equation}
where the Pantheon+ term is included only for the CC+DESI+PP
combination.

For the correlated Cosmic Chronometer measurements, we use \citep{Kavya:2025vsj}
\begin{equation}
\chi^2_{\rm CC,corr}
=
\Delta\boldsymbol{H}_{\rm c}^{\,T}
C_{\rm CC}^{-1}
\Delta\boldsymbol{H}_{\rm c},
\end{equation}
where
$
\Delta\boldsymbol{H}_{\rm c}
=
\boldsymbol{H}_{\rm th,c}
-
\boldsymbol{H}_{\rm obs,c},
$
with $\boldsymbol{H}_{\rm th,c}$ and
$\boldsymbol{H}_{\rm obs,c}$ denoting the theoretical and observed
Hubble-parameter vectors for the correlated CC measurements,
respectively, and $C_{\rm CC}$ the corresponding covariance matrix. The uncorrelated measurements contribute
\begin{equation}
\chi^2_{\rm CC,uncorr}
=
\sum_i
\frac{
\left[H_{\rm th}(z_i)-H_{{\rm obs},i}\right]^2
}{
\sigma_{H,i}^2
}.
\end{equation}
Here, $H_{\rm th}(z_i)$ denotes the theoretical Hubble parameter
predicted by the cosmological model at the redshift of the $i$th
measurement. Thus,
\begin{equation}
\chi^2_{\rm CC}
=
\chi^2_{\rm CC,corr}
+
\chi^2_{\rm CC,uncorr}.
\end{equation}
The correlated CC covariance matrix includes the measurement
uncertainties and the correlated systematic contributions adopted in
the compilation, while the uncorrelated measurements are treated with
their individual quoted uncertainties.

For DESI DR2, the model vector is constructed from the transverse and
radial BAO observables,
\begin{equation}
\boldsymbol{d}_{\rm DESI}
=
\left\{
\frac{D_M(z_i)}{r_d},
\frac{D_H(z_i)}{r_d}
\right\},
\end{equation}
and the likelihood is
\begin{equation}
\chi^2_{\rm DESI}
=
\Delta\boldsymbol{d}_{\rm DESI}^{\,T}
C_{\rm DESI}^{-1}
\Delta\boldsymbol{d}_{\rm DESI},
\end{equation}
where $
\Delta\boldsymbol{d}_{\rm DESI}
=
\boldsymbol{d}_{\rm th,DESI}
-
\boldsymbol{d}_{\rm obs,DESI}
$ The full covariance matrix supplied with the DESI DR2
measurements \citep{DESI:2025zgx} is used.

For PP, the likelihood is
\begin{equation}
\chi^2_{\rm PP}
=
\Delta\boldsymbol{\mu}^{\,T}
C_{\rm PP}^{-1}
\Delta\boldsymbol{\mu},
\end{equation}
where
$\Delta\boldsymbol{\mu}
=
\boldsymbol{\mu}_{\rm th}
-
\boldsymbol{\mu}_{\rm obs}$
and $C_{\rm PP}$ denotes the full statistical-plus-systematic PP covariance matrix. Here, the theoretical distance modulus is
\begin{equation}
\mu_{\rm th}(z)
=
5\log_{10}\!\left[
\frac{D_L(z)}{{\rm Mpc}}
\right]
+25,
\end{equation}
with $D_L(z)$ denoting the luminosity distance predicted by the
$\Lambda$CDM model defined above.
\subsection{Inference schemes}

The increasing volume and precision of modern cosmological observations have made parameter estimation a computationally demanding task, particularly when likelihood evaluations involve large covariance matrices, complex forward models, or expensive numerical integrations. As next-generation surveys continue to increase the dimensionality and statistical precision of cosmological datasets, there is growing interest in alternative inference strategies that can reduce computational cost while preserving the accuracy of posterior constraints. In this work, we perform a systematic comparison of three distinct approaches to Bayesian parameter estimation in the context of late-time $\Lambda$CDM cosmology: exact MCMC, GP emulation, and SBI. These methods represent three complementary paradigms of increasing computational sophistication, ranging from direct likelihood sampling to surrogate-model and likelihood-free approaches. Throughout this work, all three inference schemes are applied to identical cosmological models, priors, fixed $r_d$, and observational data vectors, enabling a direct and fair comparison of their statistical performance and computational efficiency. 

\subsubsection{Exact MCMC}

As a reference, we perform standard Markov Chain Monte Carlo sampling of the full likelihood. Within the Bayesian framework, the posterior distribution of the cosmological parameters is given by \citep{Trotta:2008qt}
\begin{equation}
p(\boldsymbol{\theta}\mid\boldsymbol{x})
=
\frac{\mathcal{L}(\boldsymbol{x}\mid\boldsymbol{\theta})
\,\pi(\boldsymbol{\theta})}
{\mathcal{Z}},
\end{equation}
where $\mathcal{L}(\boldsymbol{x}\mid\boldsymbol{\theta})$ denotes the likelihood function, $\pi(\boldsymbol{\theta})$ represents the prior distribution, and $\mathcal{Z}$ is the Bayesian evidence. MCMC explores the posterior distribution $p(\boldsymbol{\theta}\mid\boldsymbol{x})$ directly by constructing a Markov chain whose stationary distribution coincides with the target posterior. In the limit of sufficiently long chains and accurate likelihood evaluations, MCMC provides asymptotically exact posterior estimates and is therefore widely regarded as the gold standard for cosmological parameter inference (see, e.g., \citep{MCMC_review, Speagle:2019ffr}). Exact MCMC has the advantage of being statistically rigorous and model independent, requiring no approximation beyond the specification of the likelihood function and prior distributions. However, its computational cost scales directly with the number of likelihood evaluations. Consequently, analyses involving large datasets, high-dimensional parameter spaces, or computationally expensive theoretical predictions may require substantial computational resources, making exact sampling increasingly challenging for modern cosmological applications. In the present work, exact MCMC serves as the benchmark inference framework against which the performance of the GP-emulated and simulation-based approaches is assessed, both in terms of posterior accuracy and computational cost.

\subsubsection{GP emulator}

To accelerate repeated likelihood evaluations, we construct a
Gaussian-process emulator for the CC and DESI cosmological observables,
while retaining the Pantheon+ likelihood in its exact form, following the
general emulator framework commonly employed in cosmological analyses
\citep{Boruah:2022uac}. In this setting, the GP is trained on a set of parameter points and corresponding model predictions, and then used to interpolate the mapping $\boldsymbol{\theta}\mapsto \boldsymbol{y}(\boldsymbol{\theta})$ at negligible cost during MCMC sampling (see, e.g., \citep{GP_emulator_cosmo2,Lawrence:2017ost,Bocquet:2020tes}). The resulting GP-assisted MCMC retains the familiar workflow of standard chains while reducing the runtime by avoiding repeated calls to the expensive forward model. The accuracy of this approach depends on the density and coverage of the training set and on how well the GP kernel captures the smoothness of the target function. Gaussian processes model the target function as a distribution over functions,

\begin{equation}
f(\boldsymbol{\theta})
\sim
\mathcal{GP}
\left[
m(\boldsymbol{\theta}),
k(\boldsymbol{\theta},\boldsymbol{\theta}')
\right],
\end{equation}

where $m(\boldsymbol{\theta})$ and
$k(\boldsymbol{\theta},\boldsymbol{\theta}')$
denote the mean and covariance functions, respectively.
The kernel function encodes correlations between different points in parameter space and enables interpolation of cosmological observables at previously unseen parameter values.

In practice, the construction of an accurate emulator requires a representative sampling of the parameter space and a sufficiently expressive surrogate model capable of reproducing the cosmological predictions within the statistical precision of the data. The training set is generated using Latin Hypercube Sampling (LHS) \citep{McKay1979}, which provides an efficient space-filling coverage of the parameter domain and improves emulator performance compared to purely random sampling. To further enhance computational efficiency, we first compress the high-dimensional observable vectors using Principal Component Analysis (PCA) \citep{JolliffeCadima2016}, retaining only the dominant modes that capture the vast majority of the variance in the training set. Independent GP models are then trained on the compressed principal-component coefficients rather than on the full observable vector. This PCA-GP strategy significantly reduces the dimensionality of the emulation problem while preserving the relevant cosmological information.

For the present analysis, the GP emulator is trained on $600$ parameter
points generated using Latin Hypercube Sampling over the adopted prior
volume. The resulting $46$-dimensional CC+DESI observable vector is
standardized and compressed using PCA, retaining $99.9\%$ of the total
variance; this corresponds to three principal components. Independent
Gaussian processes are then trained for each principal component using
the kernel
\begin{equation}
k(\boldsymbol{\theta},\boldsymbol{\theta}')
=
C\,{\rm RBF}(\boldsymbol{\theta},\boldsymbol{\theta}')
+
{\rm WhiteKernel},
\end{equation}
with $\alpha=10^{-10}$, normalization of the target coefficients, and no
additional kernel restarts during optimization. The Pantheon+ likelihood
is evaluated exactly and is not emulated.

Gaussian processes provide a flexible non-parametric interpolation framework in which predictions at previously unseen parameter values are obtained from correlations learned during training. An important advantage of GP emulators is that they naturally provide uncertainty estimates associated with the interpolation, allowing the quality of the surrogate model to be assessed and monitored throughout the parameter space.
The GP predictive uncertainty is retained as an emulator diagnostic; however, it is not added as an extra covariance contribution in the
production likelihood. Instead, the emulator fidelity is assessed
independently by comparing GP predictions with exact model evaluations on
independent parameter-space points and on samples drawn from the Exact
MCMC posterior. Once trained, the emulator can generate theoretical predictions several orders of magnitude faster than direct evaluations of the cosmological model, thereby reducing the computational burden of posterior exploration.

Despite these advantages, GP-based inference remains fundamentally tied to the MCMC paradigm and therefore still requires repeated likelihood evaluations during sampling. Consequently, although emulation substantially lowers the cost of each likelihood call, the overall computational expense continues to scale with the length of the Markov chains and the dimensionality of the parameter space. The method therefore occupies an intermediate position between exact MCMC and fully amortized approaches, such as simulation-based inference, offering a compromise between computational efficiency and fidelity to the original likelihood framework.

\subsubsection{Simulation-based inference}

Finally, we consider simulation-based inference based on neural density estimation. In SBI, one uses simulations $\boldsymbol{x}\sim p(\boldsymbol{x}\mid\boldsymbol{\theta})$ and draws of $\boldsymbol{\theta}\sim p(\boldsymbol{\theta})$ to train a conditional density estimator $q_\phi(\boldsymbol{\theta}\mid\boldsymbol{x})$ (for example, via sequential neural posterior estimation), which approximates the posterior without ever evaluating the likelihood explicitly (see, e.g., \citep{SBI_review,Wang:2023vej}). The objective is to learn

\begin{equation}
q_{\phi}(\boldsymbol{\theta}\mid\boldsymbol{x})
\approx
p(\boldsymbol{\theta}\mid\boldsymbol{x}),
\end{equation}

where $\phi$ denotes the parameters of the neural density estimator. Once trained, the neural posterior can be evaluated and sampled at negligible cost for any new data vector. SBI is particularly attractive when the forward model is complex but simulatable, and when one wishes to amortize the simulation cost over many analyses.

In this work, we generate a large ensemble of synthetic realizations by drawing cosmological parameters from the chosen prior distribution and propagating them through the $\Lambda$CDM forward model to obtain the corresponding observables. For each sampled parameter vector, the theoretical observables are computed and subsequently perturbed according to the corresponding observational covariance matrices, thereby generating synthetic realizations that mimic the statistical properties of the measured CC, DESI, and Pantheon+ datasets. The resulting collection of parameter-data pairs ${(\boldsymbol{\theta}_i,\boldsymbol{x}_i)}$ constitutes the training set used for posterior learning.

We employ Sequential Neural Posterior Estimation (SNPE) \citep{Papamakarios:2016rhz, Greenberg:2019xkh}, implemented within the \texttt{sbi} framework, to learn a flexible approximation to the posterior distribution. The neural density estimator is trained directly on the simulated parameter-observation pairs and therefore learns the inverse mapping from observables to cosmological parameters. Unlike conventional likelihood-based approaches, the method does not require repeated evaluations of the likelihood function during inference. Instead, the computational effort is concentrated in a one-time training stage, after which posterior samples can be generated rapidly for a given observed dataset. Consequently, the cost of posterior evaluation becomes largely independent of the complexity of the likelihood function once training has been completed.

A key feature of this approach is the separation between training and inference costs. While the generation of simulations and neural-network training may require a non-negligible upfront investment, these costs are incurred only once. Subsequent posterior evaluations can be performed in a matter of seconds and are effectively independent of the dimensionality of the likelihood evaluation itself. This amortized inference paradigm is particularly appealing for modern cosmological analyses involving high-dimensional data vectors, large covariance matrices, or computationally expensive theoretical predictions.

The accuracy of SBI depends on the quality and coverage of the simulations used during training, the expressiveness of the neural density estimator, and the extent to which the training set adequately samples the posterior support. Nevertheless, when properly calibrated, SBI provides a powerful alternative to conventional sampling techniques, offering substantial computational savings while maintaining accurate posterior reconstruction. In the present work, we assess its performance by directly comparing the inferred cosmological constraints and computational costs against those obtained from exact MCMC and GP-assisted MCMC analyses.

Taken together, these three approaches represent progressively different strategies for Bayesian inference in cosmology. Their comparison provides a controlled framework for assessing the trade-off between computational efficiency and statistical accuracy, with exact MCMC serving as the reference benchmark against which the GP-emulated and SBI approaches are evaluated.

\paragraph*{\textbf{Implementation details:}}

All analyses were performed in Python. Exact and GP-assisted MCMC sampling were carried out using the \texttt{emcee} ensemble sampler \citep{Foreman-Mackey:2012any}. Gaussian-process emulation was implemented using the \texttt{scikit-learn} machine-learning library \citep{Pedregosa:2011ork}, while dimensionality reduction was performed through PCA using the same framework. Simulation-based inference was implemented using the \texttt{sbi} package \citep{Tejero-Cantero:2020wom}, employing SNPE with neural density estimators based on \texttt{PyTorch} \citep{Paszke:2019xhz}. Posterior visualization and statistical summaries were generated using \texttt{GetDist} \citep{Lewis:2019xzd}. Numerical computations were performed using the scientific Python ecosystem, including \texttt{NumPy} \citep{vanderWalt:2011bqk}, \texttt{SciPy} \citep{Virtanen:2019joe}, and \texttt{Matplotlib} \citep{Hunter:2007ouj}.

\section{Results and Discussion}
\label{sec:results}
\begin{figure}
    \centering
    \includegraphics[width=0.55\linewidth]{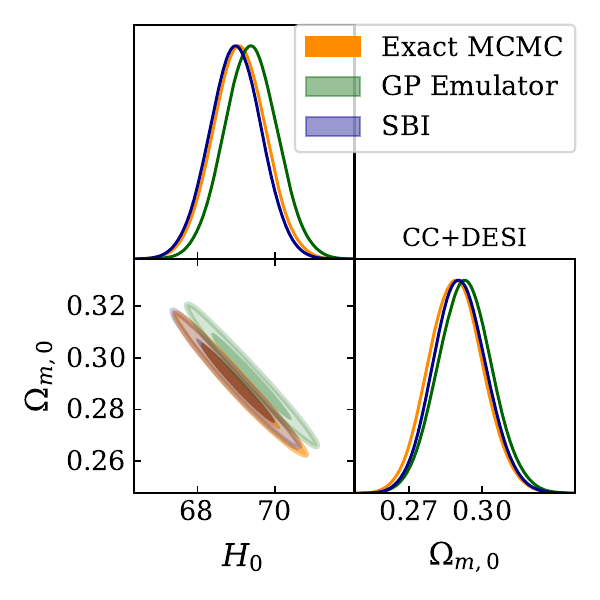}
    \caption{Posterior constraints on $H_0$ and $\Omega_{m,0}$ for the CC+DESI combination, comparing Exact MCMC (orange), the GP emulator (green), and SBI (blue).}
    \label{fig:contourcd}
\end{figure}
\begin{figure}
    \centering
    \includegraphics[width=0.85\linewidth]{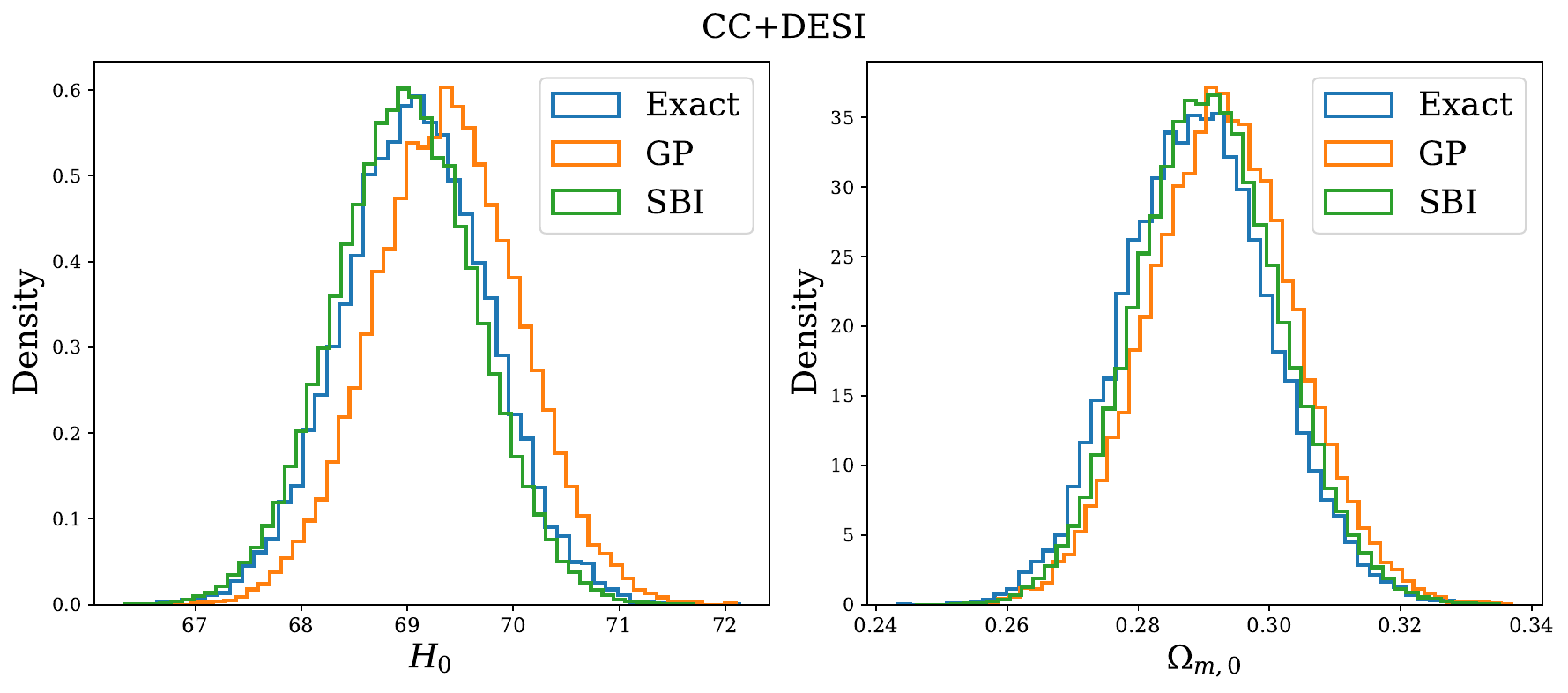}
    \caption{One-dimensional posterior distributions for $H_0$ (left) and $\Omega_{m,0}$ (right) from the three methods for the CC+DESI combination.}
    \label{fig:densitycd}
\end{figure}
\begin{figure}
    \centering
    \includegraphics[width=0.49\linewidth]{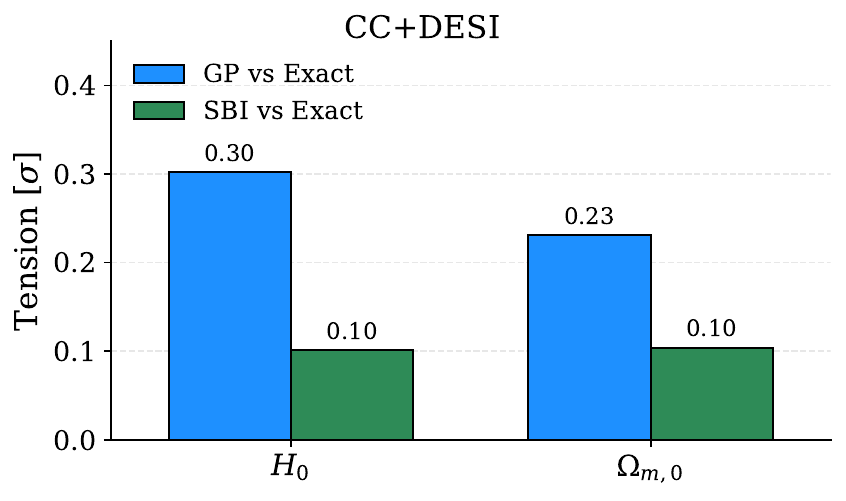}
    \includegraphics[width=0.49\linewidth]{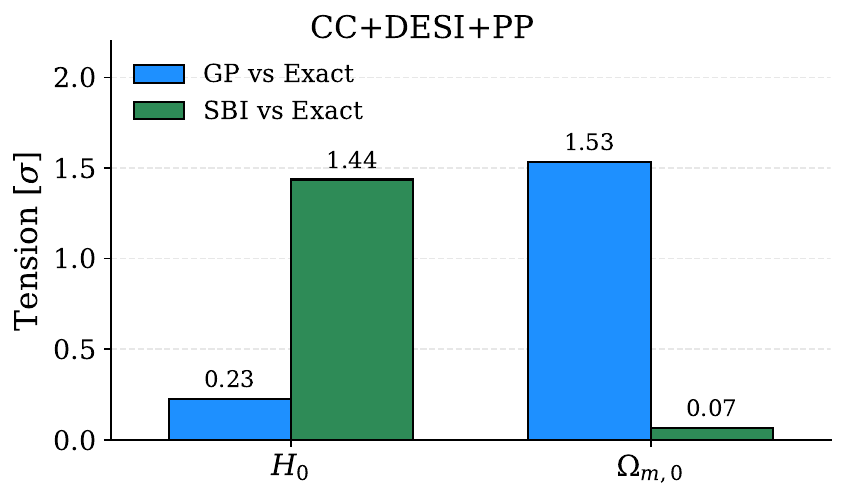}
    \caption{One-dimensional tension between approximate methods and Exact MCMC for each parameter and data combination. Left: CC+DESI tensions for GP vs.\ Exact and SBI vs.\ Exact in units of $\sigma$. Right: same for CC+DESI+PP.}
    \label{fig:tension}
\end{figure}

\begin{figure}
    \centering
    \includegraphics[width=0.55\linewidth]{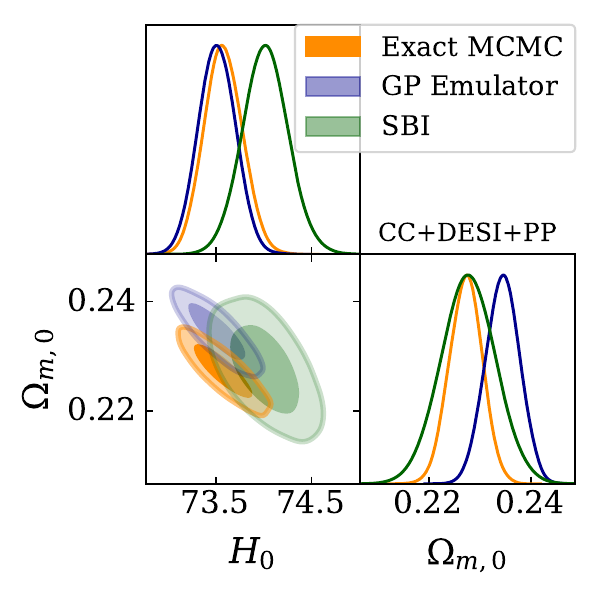}
    \caption{Posterior constraints on $H_0$ and $\Omega_{m,0}$ for the CC+DESI+PP combination.}
    \label{fig:concdpp}
\end{figure}
\begin{figure}
    \centering
    \includegraphics[width=0.85\linewidth]{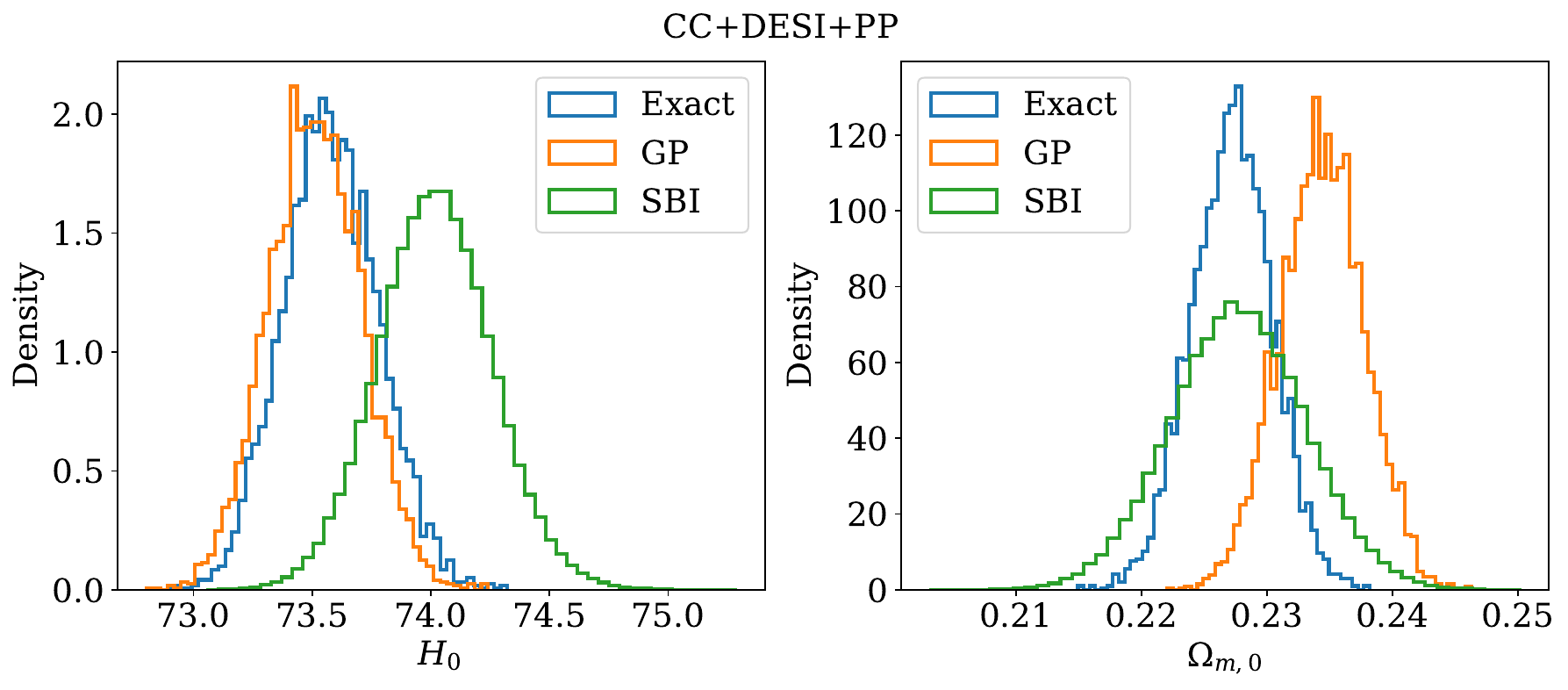}
    \caption{One-dimensional posterior distributions for $H_0$ (left) and $\Omega_{m,0}$ (right) from the three methods for the CC+DESI+PP combination.}
    \label{fig:densitypp}
\end{figure}
\begin{figure}
    \centering
    \includegraphics[width=0.49\linewidth]{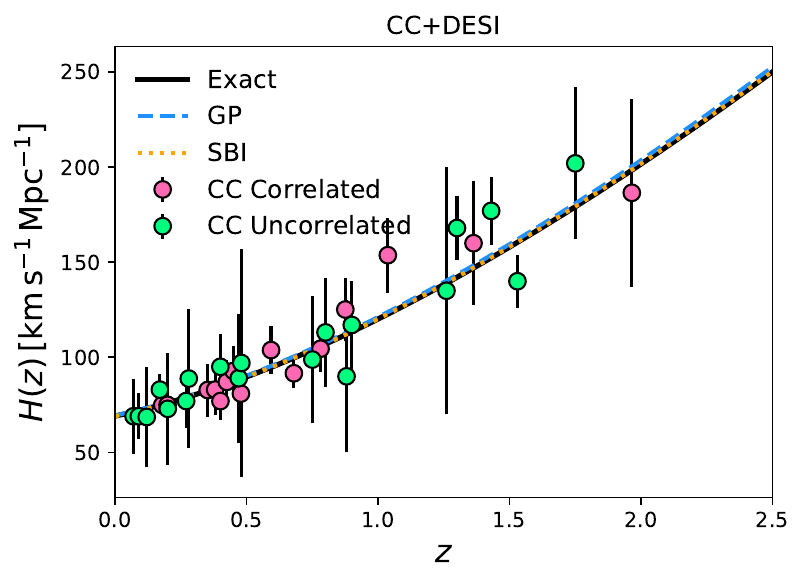}
    \includegraphics[width=0.49\linewidth]{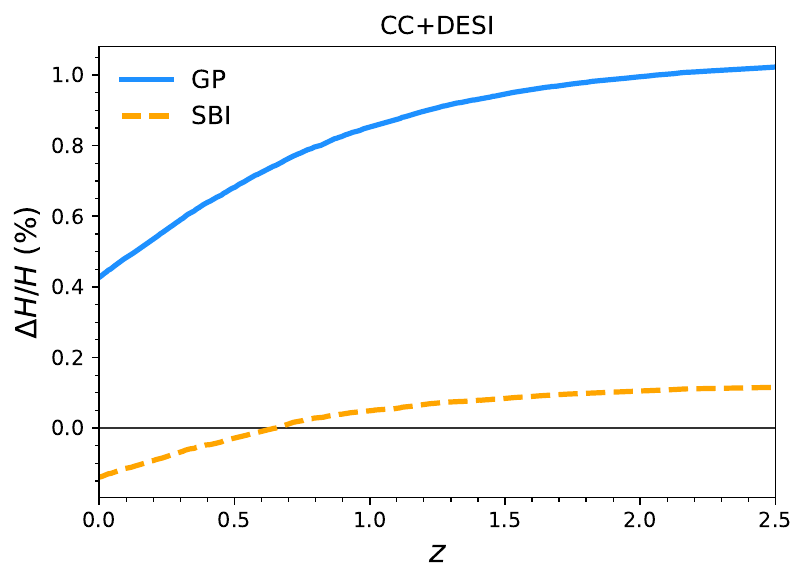}
    \includegraphics[width=0.49\linewidth]{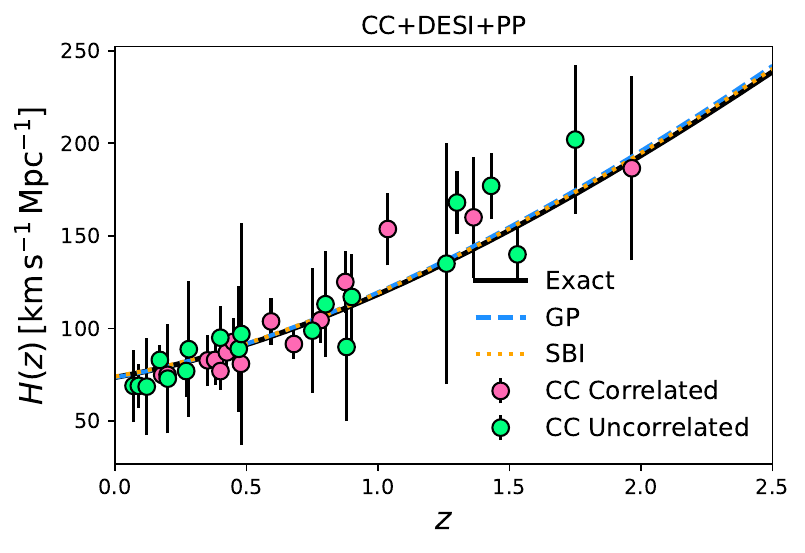}
    \includegraphics[width=0.49\linewidth]{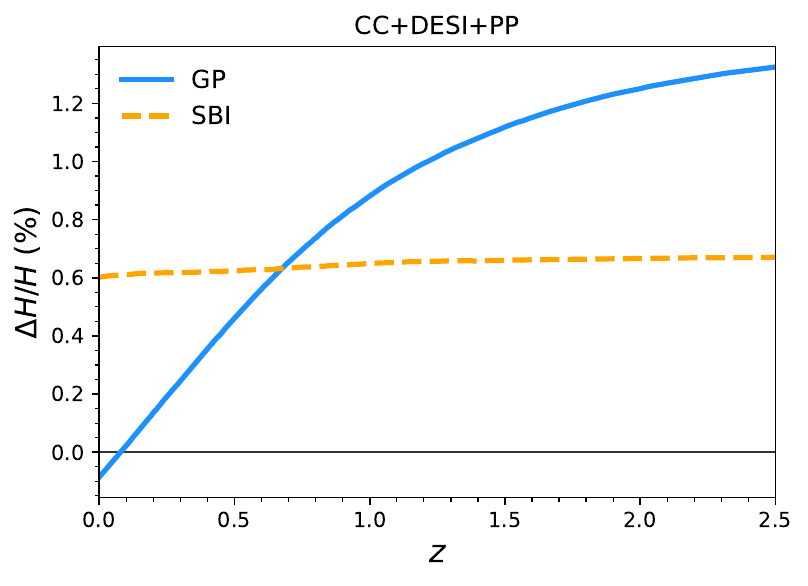}
    \caption{Reconstructed expansion histories and residual comparisons for the CC+DESI and CC+DESI+PP dataset combinations. The left panels show the median reconstructed Hubble parameter $H(z)$ obtained from Exact MCMC, GP-assisted MCMC, and SBI, together with the correlated and uncorrelated Cosmic Chronometer measurements. The right panels display the relative differences with respect to the Exact MCMC reconstruction, defined as $\Delta H/H (\%)= 100\times(H_{\rm method}-H_{\rm Exact})/H_{\rm Exact}$. For both dataset combinations, the reconstructed expansion histories inferred by the three approaches are nearly indistinguishable across the entire redshift range. The maximum deviations remain below approximately
$\sim1.3\%$ for the GP emulator and below $\sim0.6\%$ for SBI
across both dataset combinations, demonstrating that both accelerated inference methods accurately reproduce the expansion history obtained from exact likelihood sampling.}
    \label{fig:Hreconstrcuted}
\end{figure}
\begin{figure}
    \centering
    \includegraphics[width=0.95\linewidth]{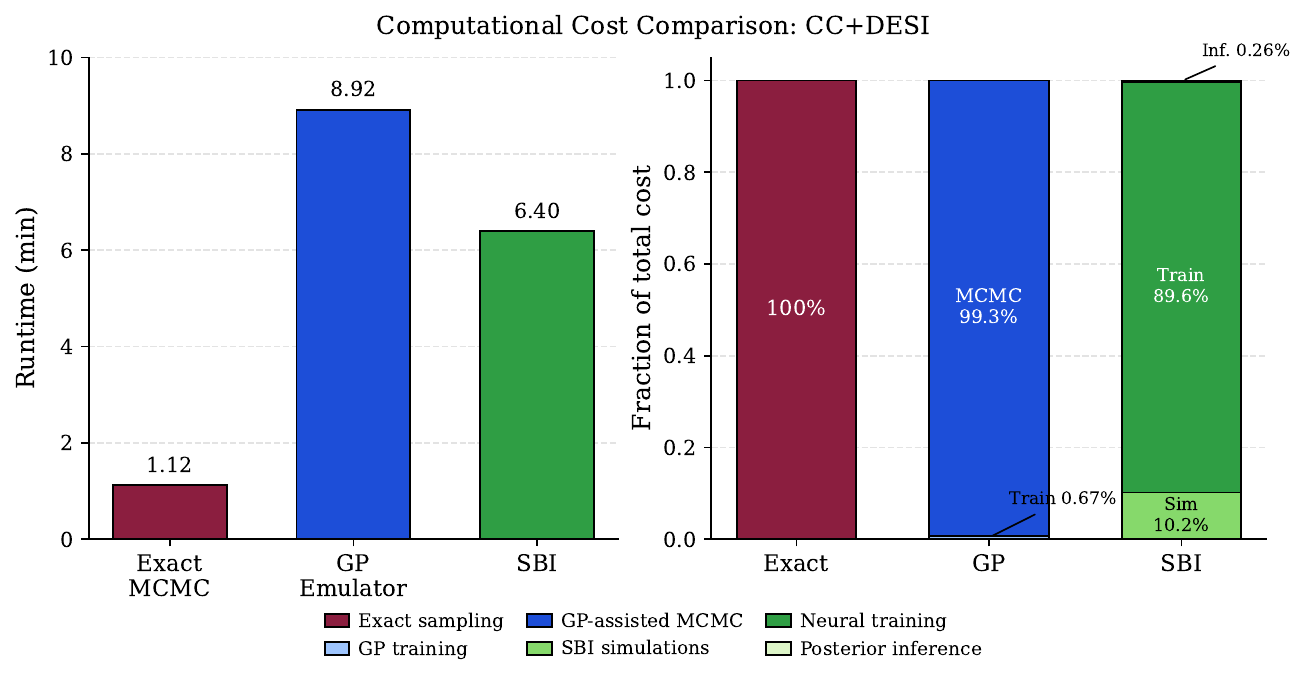}
    \caption{Computational cost comparison for the CC+DESI combination. Left: wall-clock runtime for Exact MCMC, the GP emulator (including training and GP-assisted MCMC), and SBI (including simulations, neural training, and posterior sampling). Right: decomposition of each method's total cost into simulation, training, and inference fractions. The runtimes correspond to production runs using 48 walkers and 5000 sampling steps for the likelihood-based methods and 50,000 training simulations with 100,000 posterior samples for SBI.}
    \label{fig:runtimecd}
\end{figure}
\begin{figure}
    \centering
    \includegraphics[width=0.95\linewidth]{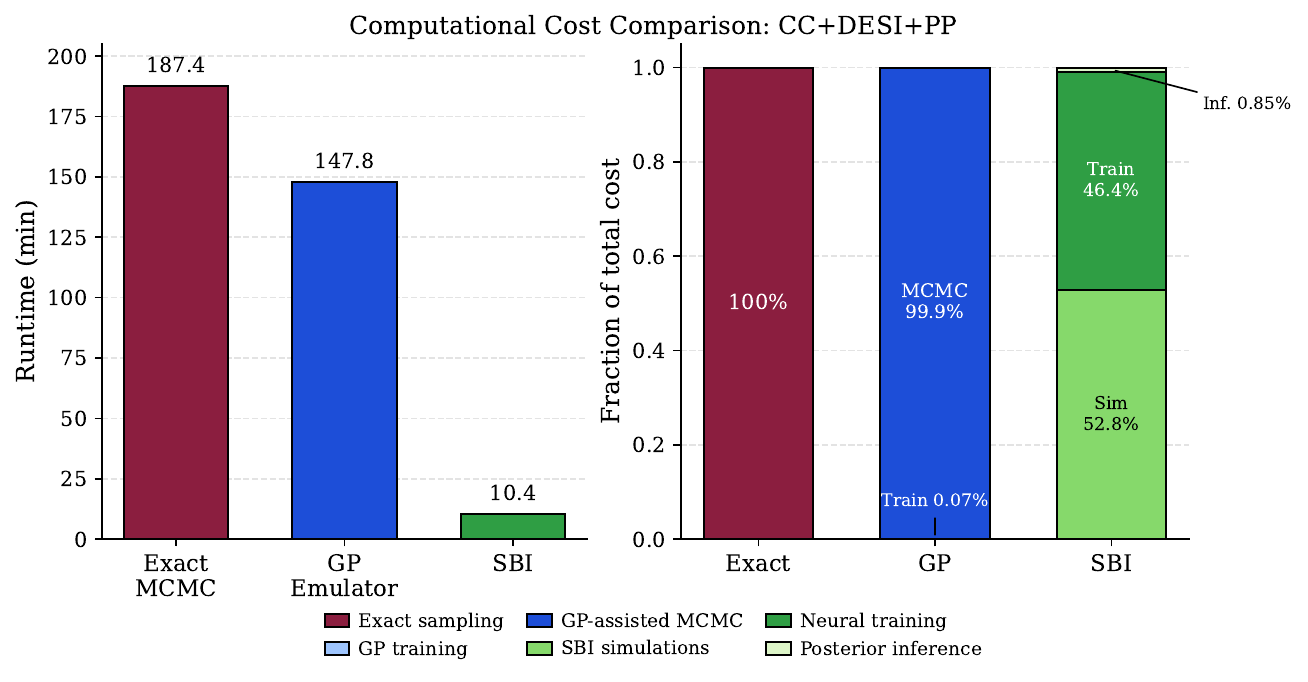}
    \caption{Computational cost comparison for the CC+DESI+PP combination. Left: total runtime for Exact MCMC, the GP emulator, and SBI. Right: fractional breakdown of the total cost into exact sampling, GP-assisted MCMC, GP and SBI training, and SBI simulations and inference. The runtimes correspond to production runs using 48 walkers and 1000 sampling steps for the likelihood-based methods and 50,000 training simulations with 100,000 posterior samples for SBI.}
    \label{fig:runtimepp}
\end{figure}

\begin{figure}
    \centering
    \includegraphics[width=0.49\linewidth]{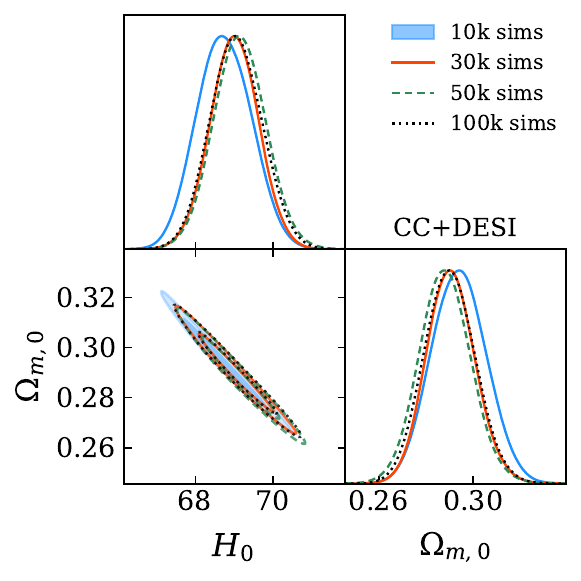}
    \includegraphics[width=0.49\linewidth]{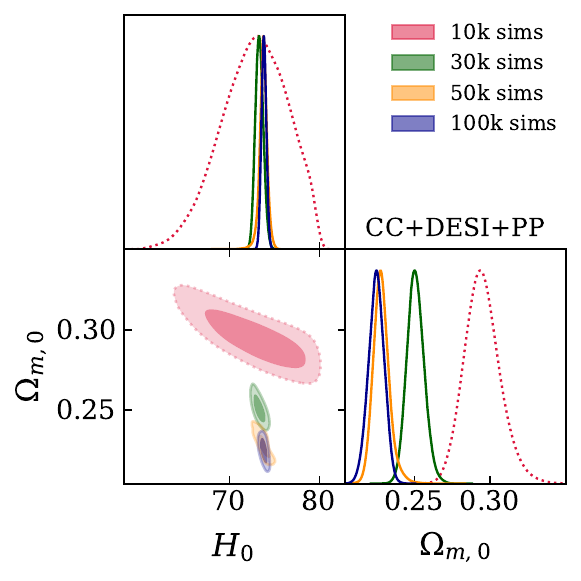}
    \caption{Convergence of SBI posteriors with the number of simulations for both data combinations. Left: CC+DESI posteriors for $H_0$ and $\Omega_{m,0}$ obtained using 10k, 30k, 50k, and 100k simulations. Right: same for CC+DESI+PP.}
    \label{fig:conv}
\end{figure}

\begin{table}[t]
\centering
\caption{Comparison of $\Lambda$CDM parameter constraints from Exact MCMC, GP emulator, and SBI for two late-time data combinations. Quoted values are posterior medians with 68\% credible intervals. The one-dimensional tension between each approximate method and Exact is defined as $T_{1\mathrm{D}} = |\Delta\mu| / \sqrt{\sigma_{\rm Exact}^2 + \sigma_{\rm method}^2}$, where $\mu$ and $\sigma$ are the posterior mean and standard deviation, respectively.}
\label{tab:method_comparison}
\begin{tabular}{llccccc}
\hline\hline
Dataset & Parameter & Exact MCMC & GP & $T_{1\mathrm{D}}$(GP) & SBI & $T_{1\mathrm{D}}$(SBI) \\
\hline
CC + DESI
& $H_0$
& $69.0838^{+0.6839}_{-0.6791}$
& $69.3788^{+0.6863}_{-0.6856}$ & $0.30\,\sigma$
& $68.9879^{+0.6668}_{-0.6711}$ & $0.10\,\sigma$
\\[1.2ex]

& $\Omega_{m,0}$
& $0.2891^{+0.0112}_{-0.0110}$
& $0.2928^{+0.0111}_{-0.0110}$ & $0.23\,\sigma$
& $0.2907^{+0.0110}_{-0.0106}$ & $0.10\,\sigma$
\\[1.2ex]\hline
CC + DESI + PP
& $H_0$
& $73.5750^{+0.2057}_{-0.1922}$
& $73.5120^{+0.1961}_{-0.1935}$ & $0.23\,\sigma$
& $74.0183^{+0.2366}_{-0.2353}$ & $1.44\,\sigma$
\\[1.2ex]

& $\Omega_{m,0}$
& $0.2273^{+0.0031}_{-0.0033}$
& $0.2344^{+0.0033}_{-0.0034}$ & $1.53\,\sigma$
& $0.2277^{+0.0053}_{-0.0053}$ & $0.07\,\sigma$
\\
\hline\hline
\end{tabular}
\end{table}

Modern cosmological analyses increasingly lean on fast surrogates and machine-learned posteriors, but their performance is rarely assessed on equal footing against fully converged MCMC. In this section, we place Exact MCMC, a GP emulator, and SBI side by side on late-time data sets, and ask a simple question: \emph{how much accuracy do we give up for a given gain in speed?}
We compare $\Lambda$CDM parameter constraints and computational costs obtained with the three inference strategies based on neural density estimation.

\subsection{Constraints from CC+DESI}

Figure~\ref{fig:contourcd} shows the joint posterior constraints on $H_0$ and $\Omega_{m,0}$ for the CC+DESI combination. The GP emulator and SBI posteriors closely overlap the Exact MCMC contours, and the one-dimensional marginals along both axes are visually indistinguishable at the level of the line thickness. This agreement is quantified in Table~\ref{tab:method_comparison}: for CC+DESI, the GP and SBI medians for $H_0$ and $\Omega_{m,0}$ differ from the Exact MCMC medians by less than $0.3\sigma$ in all cases. 

The one-dimensional histograms in Figure~\ref{fig:densitycd} make these small differences more transparent. For both $H_0$ and $\Omega_{m,0}$, the GP and SBI curves lie on top of the Exact MCMC distribution, with only mild shifts in the central values and almost identical widths. The mean differences computed from the chains are $\Delta H_0 \simeq 0.29$ and $0.10$~km\,s$^{-1}$\,Mpc$^{-1}$ for GP and SBI, respectively, and $\Delta\Omega_{m,0} \simeq 0.0037$ and $0.0016$. These shifts correspond to the tensions shown in the left panel of Figure~\ref{fig:tension}: $T_{1\mathrm{D}} \leq 0.3\sigma$ for both parameters and both approximate methods, indicating excellent agreement with Exact MCMC within the statistical precision of this dataset.

\subsection{Constraints from CC+DESI+PP}

Including the Pantheon+ supernova sample substantially increases the constraining power on late-time $\Lambda$CDM parameters. Figure~\ref{fig:concdpp} displays the corresponding joint posteriors in the $(H_0,\Omega_{m,0})$ plane for CC+DESI+PP. In this case, systematic differences between methods become visible: the GP emulator contours shift towards higher $\Omega_{m,0}$ at nearly fixed $H_0$, while the SBI contours shift towards higher $H_0$ at nearly fixed $\Omega_{m,0}$ compared to Exact MCMC.

The one-dimensional posteriors in Figure~\ref{fig:densitypp} highlight these shifts. Relative to Exact, the GP posterior for $\Omega_{m,0}$ peaks at a noticeably larger value, and the SBI posterior for $H_0$ peaks at a higher value and is somewhat broader. Table~\ref{tab:method_comparison} and the right panel of Figure~\ref{fig:tension} summarize these differences quantitatively: for CC+DESI+PP we find a tension of $T_{1\mathrm{D}} \simeq 1.53\sigma$ between the GP and Exact constraints on $\Omega_{m,0}$, and $T_{1\mathrm{D}} \simeq 1.44\sigma$ between the SBI and Exact constraints on $H_0$. The remaining parameters stay in good agreement, with tensions well below $0.3\sigma$.

\subsection{Reconstructed expansion history}
To further assess the physical consistency of the inferred cosmological constraints, we reconstruct the Hubble expansion history $H(z)$ using the median posterior parameters obtained from Exact MCMC, GP-assisted MCMC, and SBI. The resulting reconstructions for the CC+DESI and CC+DESI+PP dataset combinations are shown in Figure~\ref{fig:Hreconstrcuted}, together with the Cosmic Chronometer measurements used in the analysis.

The left panels of Figure~\ref{fig:Hreconstrcuted} demonstrate that all three inference schemes yield nearly identical expansion histories across the entire redshift range probed by the data. The reconstructed curves closely follow the observed CC measurements and remain visually indistinguishable even for the more constraining CC+DESI+PP combination. This agreement provides an important validation beyond parameter-level comparisons, showing that the accelerated inference methods recover the same underlying cosmological evolution inferred from exact likelihood sampling.

To quantify any differences, the right panels show the relative deviations with respect to the Exact MCMC reconstruction. For the CC+DESI analysis, the maximum deviations are approximately $1.02\%$ for the GP emulator and $0.14\%$ for SBI. For the more constraining CC+DESI+PP combination, the corresponding deviations are approximately $1.32\%$ and $0.67\%$, respectively. In all cases, the discrepancies remain at the percent level or below throughout the full redshift range $\left(z\in[0,2.5]\right)$. These results indicate that both GP-assisted MCMC and SBI accurately reproduce the expansion history inferred from the exact posterior, with SBI exhibiting the closest overall agreement in the present analysis.

\subsection{Computational cost and efficiency}

The reported runtimes correspond to the production settings adopted in this work. For the CC+DESI analysis, both Exact MCMC and GP-assisted MCMC were performed using 48 walkers and 5000 sampling steps, while for the computationally more demanding CC+DESI+PP combination, 48 walkers and 1000 sampling steps were employed. All runtimes were measured on the same computational environment and should therefore be interpreted as relative performance indicators rather than absolute hardware-independent benchmarks. In the SBI framework, the neural density estimator was trained using 50,000 simulated realizations, and posterior constraints were subsequently obtained from 100,000 samples drawn from the learned posterior distribution. Figures~\ref{fig:runtimecd} and \ref{fig:runtimepp} compare the computational cost of the three methods for the CC+DESI and CC+DESI+PP combinations, respectively. For CC+DESI (Figure~\ref{fig:runtimecd}), Exact MCMC completes in $1.12$~minutes, while the GP-based approach requires $8.92$ minutes once GP training and GP-assisted sampling are included. SBI lies in between, dominated by the cost of generating simulations and training the neural density estimator. The right panel of Figure~\ref{fig:runtimecd} shows the fractional runtime contributions from the different stages of each method. For the likelihood-based approaches, the computational cost is dominated by posterior sampling, whereas for SBI the majority of the runtime is associated with simulation generation and neural-network training.

For CC+DESI+PP (Figure~\ref{fig:runtimepp}) the picture changes dramatically. The Exact MCMC chain now requires $\sim 3$~hours of wall-clock runtime, while the GP-based analysis takes $\sim 2.5$~hours including emulator training and GP-assisted MCMC. In contrast, SBI produces posterior constraints that remain broadly consistent with those obtained from the likelihood-based approaches while requiring only $\sim10$ minutes of wall-clock time. The fractional breakdown in the right panel of Figure~\ref{fig:runtimepp} shows that for SBI, the dominant cost is simulations, followed by neural training, whereas the actual posterior sampling is essentially negligible. These results illustrate that SBI can provide order-of-magnitude reductions in computational cost for expensive late-time cosmological analyses while maintaining posterior constraints that are broadly consistent with those obtained from conventional likelihood-based methods. Any potential loss of accuracy associated with the approximation is assessed through a direct comparison of the recovered posterior distributions and parameter constraints.

\subsection{Convergence of simulation-based inference}

To assess the robustness of SBI as a function of simulation budget, Figure~\ref{fig:conv} shows the SBI posteriors obtained with 10k, 30k, 50k, and 100k simulations for both data combinations. For CC+DESI (left panel), the SBI constraints converge rapidly: the 30k, 50k, and 100k posteriors lie on top of each other and are statistically consistent with the Exact MCMC contours in Figure~\ref{fig:contourcd}. For CC+DESI+PP (right panel), the 10k and 30k runs still show noticeable shifts in $(H_0,\Omega_{m,0})$, while the 50k and 100k simulations yield posteriors that are stable at the level of the line thickness.

When combined with the tensions in Figure~\ref{fig:tension} and the runtimes in Figures~\ref{fig:runtimecd} and \ref{fig:runtimepp}, these convergence tests provide a practical recommendation: for CC+DESI-like data sets, $\mathcal{O}(3\times 10^4)$ simulations are sufficient for SBI to match Exact MCMC within $\lesssim 0.3\sigma$, whereas more informative CC+DESI+PP analyses require larger simulation budgets (of order $5\times 10^4$-$10^5$) to control biases in $H_0$ at the $\lesssim 1\sigma$ level.

\subsection{Validation of the inference methods}
\label{sec:validation}

To further assess the reliability of the three inference strategies,
we perform additional validation tests beyond the SBI
convergence study. First, the Exact MCMC chains are examined using
integrated autocorrelation times and effective sample sizes. Second,
the GP emulator is validated against independent test points generated
within the adopted prior volume. Finally, the calibration and coverage
properties of the SBI posterior are assessed using independent simulated
datasets. These tests provide complementary diagnostics of chain
convergence, emulator fidelity, and posterior calibration.

\begin{figure}
    \centering
    \includegraphics[width=0.7\linewidth]{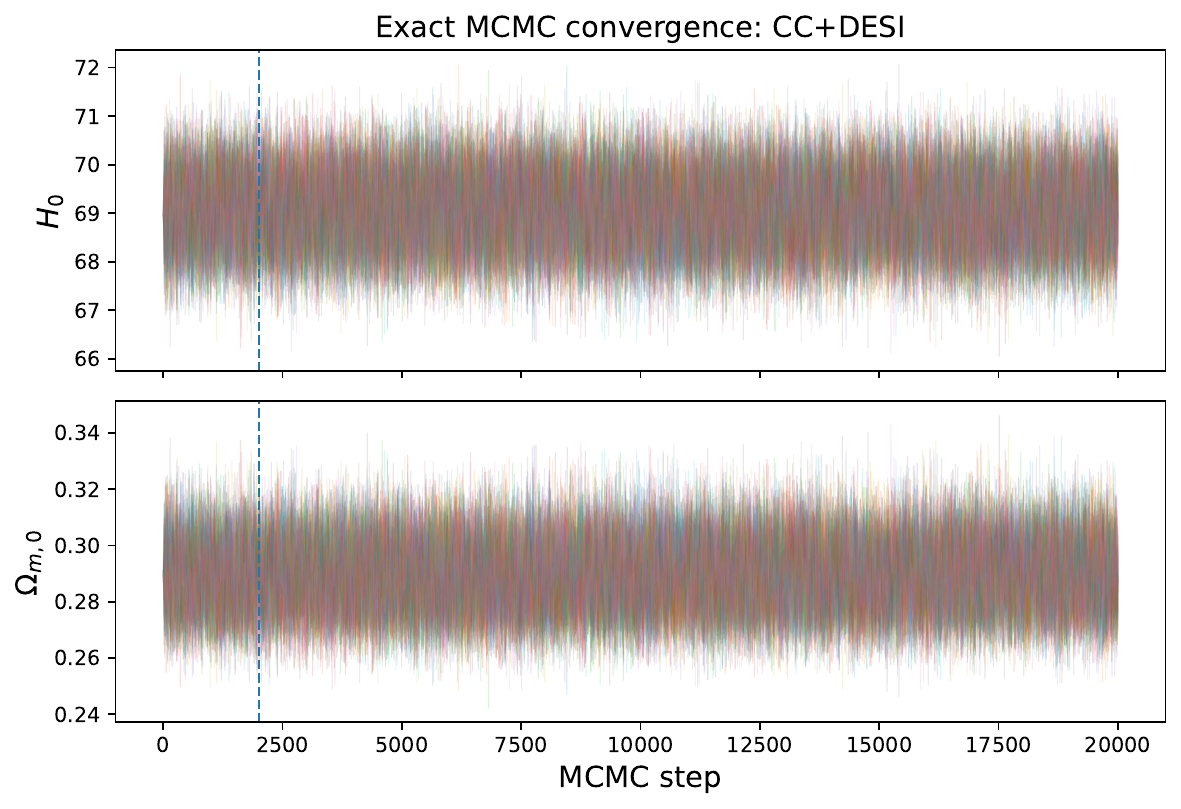}
    \includegraphics[width=0.7\linewidth]{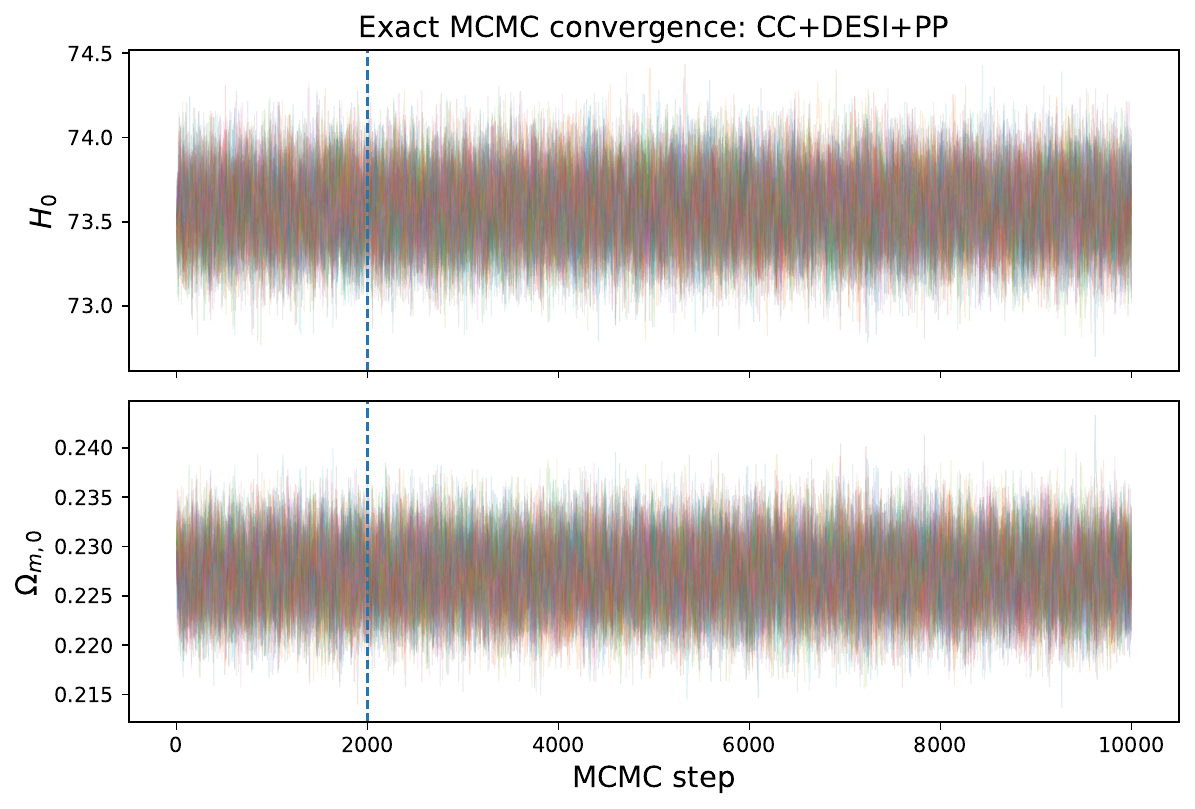}
    \caption{Trace plots of the Exact MCMC chains for the cosmological parameters $H_0$ and $\Omega_{m,0}$. The upper panel shows the CC+DESI chain, while the lower panel shows the CC+DESI+PP chain. The stationary and well-mixed behavior of the walkers provides a visual diagnostic of chain convergence.}
    \label{fig:trace}
\end{figure}

For the Exact MCMC chains, Table~\ref{tab:mcmc_convergence} summarizes
the integrated autocorrelation time $\tau_{\rm int}$, the corresponding
effective sample size $N_{\rm eff}$, the ratio of the post-burn-in
chain length to the autocorrelation time, and the mean acceptance
fraction. For both dataset combinations, the chains extend over several
hundred autocorrelation times, providing a robust basis for the comparison with the approximate inference methods.
The corresponding MCMC trace plots for both dataset combinations are also shown in Figure~\ref{fig:trace}, providing a visual check of the stationarity and mixing of the chains.

\begin{figure}
    \centering
    \includegraphics[width=0.95\linewidth]{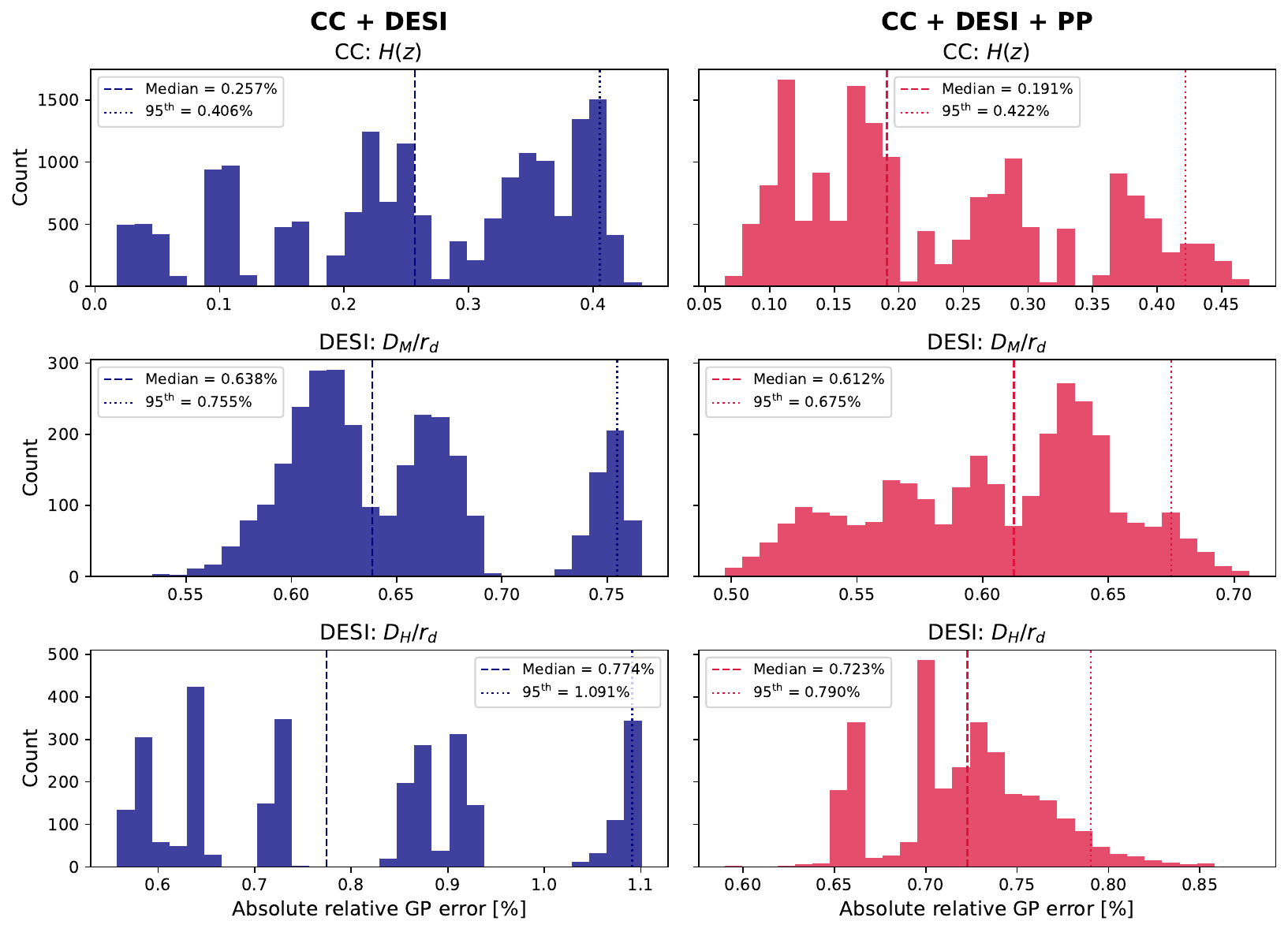}
    \caption{Relative reconstruction errors of the production GP emulator
evaluated at 500 points sampled from the Exact MCMC posterior regions.
The left column corresponds to CC+DESI and the right column to
CC+DESI+PP. From top to bottom, the panels show the errors in the CC
$H(z)$, DESI $D_M/r_d$, and DESI $D_H/r_d$ observables. 
The GP reconstruction remains accurate throughout the posterior-supported
regions, with the 95th-percentile relative errors remaining at or below
$1.1\%$.}
\label{fig:gp_validation}
\end{figure}

To quantify the accuracy of the GP approximation in the regions relevant
for parameter inference, we evaluate the production emulator at 500
points sampled from the corresponding Exact MCMC posterior for each
dataset combination. As shown in Figure~\ref{fig:gp_validation}, the
mean relative errors remain below $0.81\%$ for all emulated observables
and both dataset combinations, while the 95th-percentile errors are
below $1.1\%$. The maximum deviations in the posterior-supported regions
are $1.10\%$ for CC+DESI and $0.88\%$ for CC+DESI+PP. These results
demonstrate that the GP emulator provides a high-fidelity approximation
to the exact CC and DESI forward predictions in the regions most relevant
for cosmological parameter inference.

\begin{figure}
    \centering
    \includegraphics[width=0.9\linewidth]{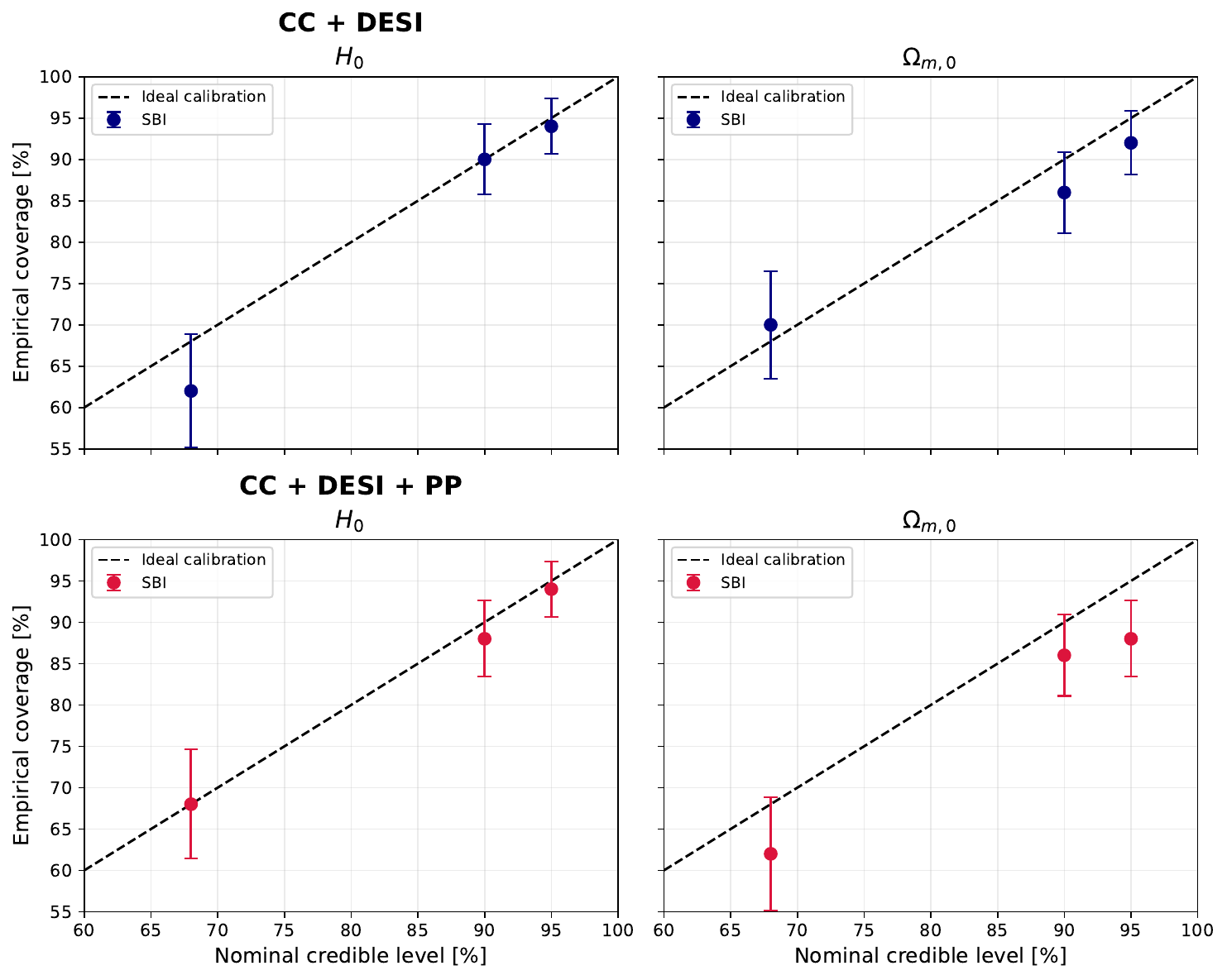}
    \caption{Empirical coverage of the SBI posterior from 50 independent
mock realizations for the two dataset combinations considered in this work.
The upper and lower rows correspond to CC+DESI and CC+DESI+PP,
respectively, while the left and right columns show $H_0$ and
$\Omega_{m,0}$. The dashed diagonal denotes ideal calibration, for which
the empirical coverage equals the nominal credible level. The error bars
represent the binomial uncertainty associated with the finite number of
calibration realizations. The measured coverages remain broadly consistent
with the nominal credible levels within these statistical uncertainties.}
\label{fig:sbi_calibration}
\end{figure}

To further assess the reliability of the SBI posterior, we perform an
independent calibration test using $50$ mock realizations for each
dataset combination. The mock data are generated using the same
cosmological prior, forward model, covariance structure, and noise model
as adopted in the production SBI analysis, while the neural posterior is
independently trained and subsequently tested on these new realizations.
We evaluate the empirical coverage of the $68\%$, $90\%$, and $95\%$
credible regions for both $H_0$ and $\Omega_{m,0}$. For CC+DESI, the
coverages for $H_0$ are $62.0\%\pm6.9\%$, $90.0\%\pm4.2\%$, and
$94.0\%\pm3.4\%$, while those for $\Omega_{m,0}$ are
$70.0\%\pm6.5\%$, $86.0\%\pm4.9\%$, and $92.0\%\pm3.8\%$ at the
respective nominal levels. For CC+DESI+PP, the corresponding coverages
are $68.0\%\pm6.6\%$, $88.0\%\pm4.6\%$, and $94.0\%\pm3.4\%$ for $H_0$,
and $62.0\%\pm6.9\%$, $86.0\%\pm4.9\%$, and $88.0\%\pm4.6\%$ for
$\Omega_{m,0}$. As shown in Figure~\ref{fig:sbi_calibration}, the
empirical coverages broadly follow the ideal calibration relation, with
the deviations remaining compatible with the finite-sample uncertainties
of the calibration experiment. This provides an additional validation of
the SBI posterior beyond the convergence tests based on increasing the
simulation budget.

\begin{table}[ht!]
\centering
\caption{Quantitative convergence diagnostics for the independent
Exact MCMC validation chains. The integrated autocorrelation time
$\tau_{\rm int}$, effective sample size $N_{\rm eff}$, ratio of the
post-burn-in chain length to $\tau_{\rm int}$, and mean acceptance fraction are reported for each cosmological parameter.}
\label{tab:mcmc_convergence}
\begin{tabular}{llrrrr}
\hline\hline
\textbf{Dataset} &
\textbf{Parameter} &
\boldmath$\tau_{\rm int}$ &
\boldmath$N_{\rm eff}$ &
\boldmath$N_{\rm post}/\tau_{\rm int}$ &
\textbf{Acceptance} \\
\hline
CC+DESI
& $H_0$
& 31.81
& 27161
& 565.9
& 0.7146 \\

CC+DESI
& $\Omega_{m,0}$
& 32.15
& 26876
& 559.9
& 0.7146 \\

CC+DESI+PP
& $H_0$
& 30.80
& 12469
& 259.8
& 0.7143 \\

CC+DESI+PP
& $\Omega_{m,0}$
& 30.70
& 12508
& 260.6
& 0.7143 \\

\hline\hline
\end{tabular}
\end{table}
\section{Summary and Future Outlook}
\label{sec:conc}

The rapid growth of modern cosmological datasets has transformed parameter inference into a computationally intensive component of cosmological analysis. As likelihood functions become increasingly expensive to evaluate and future surveys deliver ever larger and higher-dimensional datasets, the development of reliable accelerated inference techniques is becoming an essential requirement rather than a computational convenience. In this work, we performed a controlled comparison of three complementary Bayesian inference paradigms within the framework of late-time $\Lambda$CDM cosmology: Exact MCMC, GP-assisted MCMC, and SBI.
Using two representative low-redshift dataset combinations of increasing complexity, namely CC+DESI and CC+DESI+PP, we investigated the extent to which accelerated inference techniques can reproduce the statistical conclusions obtained from conventional likelihood sampling. By employing identical cosmological models, priors, and observational datasets across all analyses, the comparison isolates the impact of the inference methodology itself and provides a transparent assessment of the trade-off between computational efficiency and statistical fidelity.

Our results demonstrate that both GP emulation and SBI are capable of recovering cosmological constraints that remain broadly consistent with those obtained from exact likelihood sampling. For moderately constraining datasets, the agreement between all three approaches is excellent, while for the more informative CC+DESI+PP combination small but measurable differences emerge at the level of the posterior distributions. Nevertheless, the reconstructed expansion histories remain remarkably stable, indicating that the accelerated methods successfully recover the underlying cosmological evolution inferred from the exact posterior. These findings suggest that, at least for low-dimensional late-time cosmological analyses, the dominant cosmological information can be preserved even when significant approximations are introduced into the inference pipeline. 
The present benchmark, therefore, establishes a controlled demonstration of
posterior fidelity in a low-dimensional, smooth inference problem,
while providing a quantitative reference point for assessing more
challenging applications.

From a computational perspective, the comparison reveals that the relative
advantage of each method depends on the cost and complexity of the
underlying likelihood. Exact MCMC remains the statistically rigorous
reference, while GP emulation reduces the cost of repeated theoretical
evaluations but remains coupled to MCMC sampling and therefore does not
necessarily reduce the total runtime for comparatively inexpensive
likelihoods. SBI represents a qualitatively different paradigm in which the computational effort is shifted to an upfront simulation and training stage, after which posterior evaluation becomes essentially independent of the complexity of the likelihood. The resulting amortization of inference costs makes SBI particularly attractive for applications requiring repeated analyses, parameter forecasts, survey optimization studies, or large ensembles of mock realizations.

Beyond the specific $\Lambda$CDM application considered here, the broader implication of this work is that future cosmological inference may increasingly rely on hybrid strategies that combine physical modeling with machine-learning-based posterior estimation. In particular, the balance between computational efficiency and statistical robustness is likely to become increasingly important as upcoming surveys such as DESI \citep{DESI:2024mwx}, Euclid \citep{Euclid:2019clj}, Rubin LSST \citep{LSST:2008ijt}, and SKA \citep{SKA:2018ckk} deliver datasets whose statistical precision exceeds the practical capabilities of traditional sampling approaches.
In such environments, validated accelerated inference methods could become
increasingly valuable for timely scientific exploitation of these datasets,
provided that their accuracy and calibration are established for the
specific models and likelihoods under consideration.

Several natural extensions of the present analysis deserve further investigation. First, the comparison should be repeated in higher-dimensional cosmological models, including dynamical dark-energy scenarios \citep{Chevallier:2000qy, Linder:2002et, Copeland:2006wr, DESI:2025wyn}, modified gravity frameworks \citep{DeFelice:2010aj, Koyama:2015vza, Cai:2015emx, Kavya:2024ssu, Mishra:2025kzu}, interacting dark-sector models \citep{vanderWesthuizen:2023hcl, Kolhatkar:2026ixl}, and extensions motivated by current cosmological tensions \citep{DiValentino:2021izs}. Such analyses would provide a more stringent test of emulator performance and posterior-learning techniques in parameter spaces substantially larger than those considered here. Second, future studies should investigate the propagation of emulator uncertainties and neural-posterior calibration errors into cosmological parameter constraints, particularly in regimes where observational precision becomes comparable to the approximation error introduced by the inference method. Third, recent advances in normalizing flows, diffusion-based posterior models, active learning, and adaptive simulation strategies offer promising opportunities for further reducing the simulation budgets required by SBI while maintaining high accuracy.
These extensions will also provide a stronger test of whether the
accuracy and computational benefits demonstrated here persist as the
parameter dimensionality, nuisance-parameter content, and non-Gaussianity
of the inference problem increase.

Overall, the results presented here demonstrate that accelerated inference
techniques can provide accurate and computationally attractive alternatives
to direct likelihood sampling for the low-dimensional late-time
$\Lambda$CDM problem considered in this work.
 While exact MCMC remains the benchmark against which all approximate methods should be judged, both GP emulation and simulation-based inference provide viable pathways toward scalable cosmological inference in the era of precision cosmology.
Establishing the domains in which these methods remain reliable, and
understanding their limitations as datasets and theoretical models become
increasingly complex, will be an important step toward the next
generation of cosmological analyses.

\textbf{Data availability:} There are no new data associated with this article.\\

\section*{Acknowledgment}  
We are very grateful to the honorable referees and the editor for the illuminating suggestions that have significantly improved our work in terms of research quality and presentation.
\bibliography{main}
\bibliographystyle{aasjournal}
\nocite{}



\end{document}